\documentclass[doublecol]{epl2} 
\usepackage{float, makecell, amsmath, lineno, graphicx, subfigure}
\usepackage[hidelinks]{hyperref}
\hypersetup{colorlinks=true, allcolors=blue}
\title{Anti-correlation network among China A-shares}
\shorttitle{Anti-correlation network among China A-shares} %Insert here a short version of the title if it exceeds 70 characters

\author{Peng Liu \thanks{E-mail: \email{pengliu@xaufe.edu.cn (corresponding author)}}}%\orcidlink{0000-0002-9115-2055}
\shortauthor{P. Liu}

\institute{                    
  School of Information, Xi'an University of Finance and Economics, Xi'an 710100, Shaanxi, P.R. China
}

\abstract{
The correlation-based financial networks are studied intensively.
However, previous studies ignored the importance of the anti-correlation.
This paper is the first to consider the anti-correlation and positive correlation separately,
and accordingly construct the weighted temporal anti-correlation and positive correlation networks among stocks listed in the Shanghai and Shenzhen stock exchanges.
For both types of networks during the first 24 years of this century,
fundamental topological measurements are analyzed systematically.
This paper unveils some essential differences in these topological measurements between the anti-correlation and positive correlation networks.
It also observes an asymmetry effect between the stock market decline and rise.
The methodology proposed in this paper has the potential to reveal significant differences
in the topological structure and dynamics of a complex financial system, stock behavior, investment portfolios, and risk management,
offering insights that are not visible when all correlations are considered together.
More importantly,
this paper proposes a new direction for studying complex systems: the anti-correlation network.
It is well worth reexamining previous relevant studies using this new methodology.
}

\begin{document}
\maketitle

%%%%%%%%%%%%%%%%%%%%%%%%%%%%%%%%%%%%%%%%%%%%%%%%%%%%%%%%%%%%%%%%%
%%%%%%%%%%%%%%%%%%%%%%%%%%%%%%%%%%%%%%%%%%%%%%%%%%%%%%%%%%%%%%%%%
%%%%%%%%%%%%%%%%%%%%%%%%%%%%%%%%%%%%%%%%%%%%%%%%%%%%%%%%%%%%%%%%%
\section{Introduction}
\label{introduction}
Understanding complex systems is a long-standing and important issue.
Therefore, the Nobel Prize in Physics 2021 was awarded for groundbreaking studies on complex systems.
Since the beginning of this century,
network science has gradually become a powerful theory and tool for studying various complex systems in both nature and human society.
Taking advantage of the network approach,
scholars have gained a lot of valuable insights into various research topics from multiple academic fields~{\cite{ smallworld, network_medicine, network_ecology, network_finance, network_economy, EPL-1, EPL-2}}.
These achievements are not accessible with traditional ideas.

The financial market is a typical complex system in which market participants interact with each other nonlinearly~{\cite{FM-complex}}.
Therefore, the network approach is also widely used to understand the complex financial system.
The pioneering work in this direction was proposed by Mantegna,
who constructed networks among stocks listed in the New York Stock Exchange for the first time~{\cite{MST-1}}.
Mantegna built the connections of the networks using the correlation coefficients among the time series of fluctuations of daily logarithmic prices of stocks~{\cite{MST-1}}.
This work used a Minimal Spanning Tree (MST) to select the most important connections and found a meaningful hierarchical structure in the US stock market for the first time~{\cite{MST-1}}.
Many scholars followed this approach because of its simplicity and clear physics picture.
Consequently, MST networks in different financial markets were constructed
to investigate various aspects of the financial complex system~{\cite{MST-2, MST-3, MST-4, MST-5, MST-6, MST-7, MST-8, MST-9}}.

In addition to the MST method, researchers further used the Planar Maximally Filtered Graph (PMFG) method~{\cite{PMFG-1, PMFG-2, PMFG-3, PMFG-4}}
and the threshold method~{\cite{Threshold-1, Threshold-2, Threshold-3, Threshold-4, Threshold-5, Threshold-6}}
to construct financial networks based on the correlation coefficients among stocks.
These three methods are currently the mainstream methods for constructing correlation-based networks.
These three methods all aim to use the most important information to construct networks
and to ignore the so-called ``noise" associated with small correlation coefficients.
However, there is no criterion for distinguishing important information from the so-called ``noise".
More importantly, these three methods may completely ignore the connections associated with anti-correlation coefficients (the smaller the negative correlation coefficient a connection is associated with, the stronger the impact that connection has).
As a result,
the networks based on these three methods result in the loss of a great deal of useful information.

To extract the whole information about the correlation coefficients among stocks,
scholars also studied fully connected weighted correlation-based networks~{\cite{fully-1, fully-2, fully-3, fully-4}}.
In these fully connected weighted networks, the distance $d_{ij}$ between stocks $i$ and $j$ equals $\sqrt{2\left( 1 - \rho_{ij} \right)}$,
and the weight $w_{ij}$ of the connection between these two stocks equals $e^{-d_{ij}}$,
where $\rho_{ij}$ denotes the correlation coefficients.
From the formula of weight $w_{ij}$ it is evident that 
these studies did not consider the anti-correlations and positive correlations equally
and thus significantly ignored the connections associated with anti-correlation coefficients.

As far as we are informed,
the previous studies all ignored the importance of the connections associated with anti-correlation coefficients in correlation-based financial networks.
In those networks,
the anti-correlation and positive correlation connections
have opposite effects on the dynamics of stock price fluctuations. 
Compared with the positive correlation connections,
the anti-correlation connections play a more critical role in understanding specific properties of the complex financial systems.
For example, the anti-correlation connections can diversify market risk,
and it consequently plays a crucial role in optimizing investment portfolios and risk management.

To pay attention to the anti-correlation connections,
this paper is the first to consider the anti-correlation and positive correlation connections separately,
and accordingly construct the weighted temporal anti-correlation and positive correlation networks among stocks listed in the Shanghai and Shenzhen stock exchanges.
This work focuses on the differences in topological structures between the anti-correlation and positive correlation networks during the first 24 years of this century,
then unveils some essential differences between the anti-correlation and positive correlation networks.

\section{Data}
This paper analyzes the daily closure prices of 5,329 stocks, which are all stocks listed in the Shanghai and Shenzhen stock exchanges on Dec. 31, 2023.
There are 13,687,721 records of daily closure prices on 5,816 trading days during the 24 years from Jan. 1, 2000 to Dec. 31, 2023.

This 24-year period covers the 2001-2005 Chinese stock market slump,
the 2007-2008 global financial crisis,
the 2015-2016 Chinese stock market turbulence,
the plummeting caused by the COVID-19 pandemic outbreak in early 2020~{\cite{COVID-19-1, COVID-19-2}},
as well as other slight market crashes.
Therefore, this 24-year period especially allows us to study the dynamics of the anti-correlation and positive correlation networks under market crashes.

\section{Methodologies for network construction and analysis}
This section first introduces the methodology for constructing the anti-correlation and positive correlation networks,
and then the network analysis methodology.

Assume we have $N$ stocks.
The anti-correlation and positive correlation networks among these $N$ stocks are constructed based on the correlation coefficients~{\cite{MST-1}}.
The correlation coefficient $\rho_{ij}$ between stocks $i$ and $j$ is mathematically defined as eq.~(\ref{Eq:PCC}).
The $i$ and $j$ ($i, j = 1, 2, ..., N$) are the numerical labels of stocks.
\begin{eqnarray}
	\label{Eq:PCC}
	\rho_{ij} = \frac{\left< R_{i} R_{j} \right> - \left< R_{i} \right> \left< R_{j} \right>} {\sqrt{\left( \left< R^{2}_{i} \right> - \left< R_{i} \right>^{2} \right) \left( \left< R^{2}_{j} \right> - \left< R_{j} \right>^{2} \right)}} 
\end{eqnarray}
where $R_{i} \left( t \right) = \ln P_{i}\left(t\right) - \ln P_{i}\left(t-1\right)$ is the logarithmic return~{\cite{logreturn}},
$P_{i}\left(t\right)$ is the daily closure price of stock $i$ on trading day $t$,
and $\left< ~ \right>$ represents temporal average over a specific time window.
All correlation coefficients $\rho_{ij}$ construct a $N \times N$ symmetrical correlation matrix.

Based on this correlation matrix, we can define the weight matrices $W^{a}$ and $W$ as eqs.~(\ref{Eq:Wa}) and~(\ref{Eq:W}) for the anti-correlation and positive correlation networks, respectively.
The elements of a weight matrix represent edge weights between nodes. Here, a node represents a stock.
\begin{eqnarray}
	\label{Eq:Wa}
	W^{a}_{ij} =
	\begin{cases} 
		\left| \rho_{ij} \right|,  &  \rho_{ij} < 0\\[8pt]
		0,  &  \mbox{otherwise}
	\end{cases}
\end{eqnarray}
\begin{eqnarray}
	\label{Eq:W}
	W_{ij} =
	\begin{cases} 
		\rho_{ij},  &  \rho_{ij} > 0 \mbox{ and } i \neq j\\[8pt]
		0,  & \mbox{otherwise}
	\end{cases}
\end{eqnarray}

According to the weight matrices $W^{a}$ and $W$, we can get the binary adjacency matrices $A^{a}$ and $A$ for the anti-correlation and positive correlation networks, respectively.
$A^{a}_{ij} = 1$ if $W^{a}_{ij} > \theta$, and $A^{a}_{ij} = 0$ otherwise;
$A_{ij} = 1$ if $W_{ij} > \theta$, and $A_{ij} = 0$ otherwise.
The $\theta$ is a threshold parameter.
An element in a binary adjacency matrix determines whether an edge exists between two specific nodes.
The results reported in this paper are obtained with $\theta = 0$.
For robustness analysis,
this paper varies $\theta$ from 0 to $0.10$,
and no significant differences in the results are found.

To study the temporal evolution characteristics of the anti-correlation and positive correlation networks,
this paper constructs networks by using the sliding time window technique as described in~{\cite{fully-1}}, which is widely used in literature.
According to this technique,
the $m$th network starts on the $\left[1 + \left(m-1\right) \delta t \right]$th trading day and ends on the $\left [\left(m-1\right) \delta t + L \right]$th trading day,
where $L$ is the length of time window,
and $\delta t$ is the step that the time window slides forward.
This study sets $L$ and $\delta t$ as 26 and 15 trading days, respectively.
Such settings allow time windows to cover all trading days during the 24 years.
As a result, 387 time windows are obtained (so, $m = 1, 2, ..., 387$).
For robustness analysis,
this paper slightly varies $L$ and $\delta t$,
and no significant differences in the results are found.

In each time window, we select the stocks whose daily closure prices are available on all 26 trading days,
then calculate the correlation matrix among these stocks,
and accordingly construct the anti-correlation and positive correlation networks by removing isolated nodes.

This technique constructs 387 anti-correlation and positive correlation networks, respectively.
These 774 networks represent the temporal evolution of the Chinese stock market system over 24 years.
To quantitatively investigate the network evolution characteristics,
this paper systematically analyzes the most fundamental network topological measurements:
the node's strength,
the assortativity coefficient,
the average local clustering coefficient,
and the average shortest path length.
Here, this paper formulates these measurements for the positive correlation network only
since the same metrics are calculated for both the positive correlation and anti-correlation matrices.
In the following formulae, the $N$, $m$, $A$, and $W$ denote the number of nodes, the number of edges, the binary adjacency matrix, and the weight matrix for a network, respectively.
The $i$, $j$, and $k$ ($i, j, k = 1, 2, ..., N$) are the numerical labels of nodes.

The degree $k_{i}$ and strength $s_{i}$ of node $i$ are defined as eq.~(\ref{Eq:degree}).
These two measurements measure the importance of a node in a network~{\cite{degreeandstrength}}.
\begin{eqnarray}
	\label{Eq:degree}
	k_{i} = \sum\limits_{j}^{N} A_{ij}, \, \, \, s_{i} = \sum\limits_{j}^{N} W_{ij}
\end{eqnarray}

The assortativity coefficient $r$ by the node's strength is another important measurement for a network.
It measures the tendency of two nodes with a similar strength to be linked by an edge~{\cite{assortativity-1}}.
It is defined as eq.~(\ref{Eq:assortativity}), where $\delta_{ij}$ is the Kronecker delta function.
\begin{eqnarray}
	\label{Eq:assortativity}
	r = \frac{\sum\limits_{ij}^{N} \left( A_{ij} - k_{i}k_{j}/2m\right) s_{i}s_{j}}{\sum\limits_{ij}^{N} \left( k_{i}\delta_{ij} - k_{i}k_{j}/2m\right) s_{i}s_{j}}
\end{eqnarray}

The local clustering coefficient $C_{i}$ of node $i$ measures the occurrence of triangles attached to node $i$, which is a special case of motifs~{\cite{Threshold-2, fully-1, smallworld}}.
It is given by eq.~(\ref{Eq:clustering}).
 \begin{eqnarray}
	\label{Eq:clustering}
	C_{i} = \frac{1}{k_{i}\left(k_{i} - 1\right)}   \sum\limits_{jk}^{N}  \left(\widehat{W}_{ij}  \widehat{W}_{ik}  \widehat{W}_{jk} \right)^{1/3}
\end{eqnarray}
where $\widehat{W}_{ij}$ is the edge weight normalized by the maximum weight in a network, and
$\widehat{W}_{ij} = W_{ij}/\mbox{max}\left(W_{ij}\right)$.
The average local clustering coefficient $\left<C\right>$ is the average $C_{i}$ over all nodes.

The average shortest path length $\left< L \right>$ is a measurement to characterize the typical separation between two nodes in a network.
This measurement is important for understanding the shock propagation in a financial network.
For the network studied here, $\left< L \right>$ is given by eq.~(\ref{Eq:length}).
 \begin{eqnarray}
	\label{Eq:length}
	\left< L \right> = \frac{1}{N\left(N-1\right)}\sum\limits_{i \neq j}^{N}l_{ij}
\end{eqnarray}
where $l_{ij}$ is the shortest path length from node $i$ to node $j$.
The shortest path is a path with the minimum sum of edge distances.
The edge distance $d_{ij}$ between nodes $i$ and $j$ is defined as $d_{ij} = \sqrt{2 \left( 1 - W_{ij} \right)}$~{\cite{MST-1}}.
The measurement $\left< L \right>$ is only valid for connected networks.
Therefore, this paper only considers the largest connected component of a network when calculating $\left< L \right>$.
For the networks studied in this paper,
all positive correlation networks are connected,
while only seven anti-correlation networks are disconnected.

\section{Results and discussion}
\begin{figure}[H]
	\centering
	\includegraphics[width=1.0\linewidth]{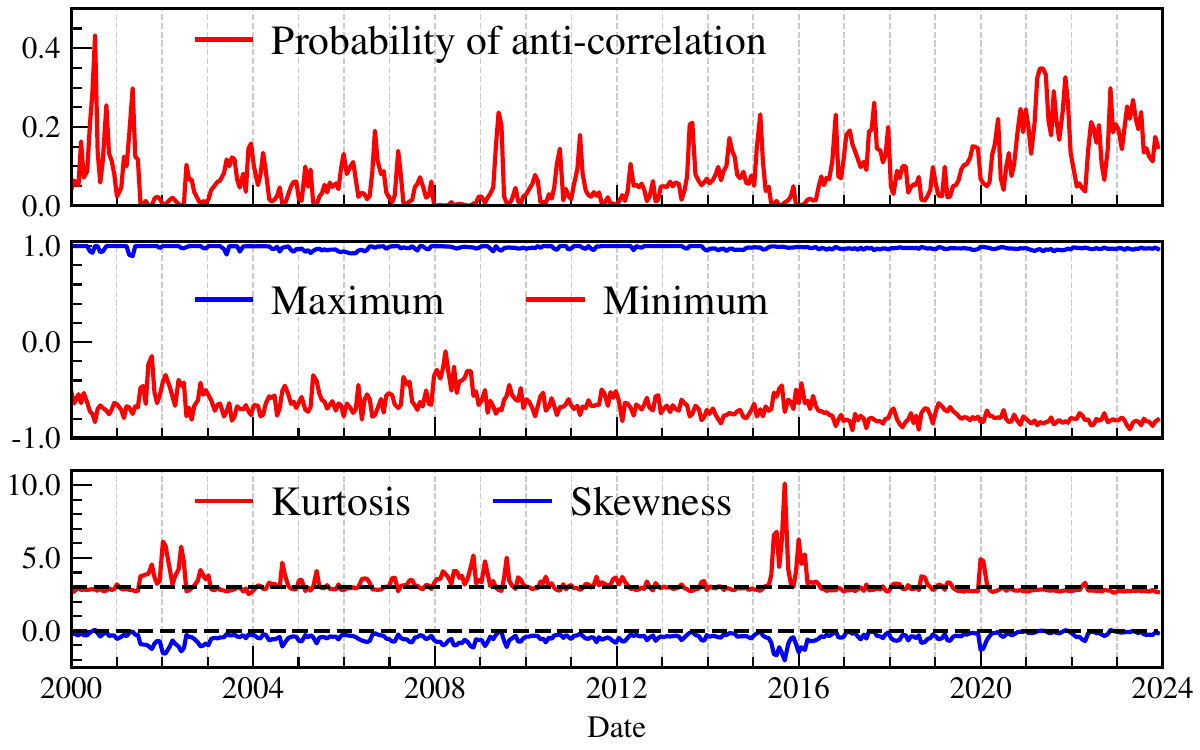}
	\caption{The distribution properties of the correlation coefficients $\rho_{ij}$ in each time window.
		The top panel presents the probability of anti-correlation $p\left( \rho_{ij} < 0 \right)$.
		The middle panel presents the maximum (upper blue curve) and minimum (lower red curve) of $\rho_{ij}$.
		The bottom panel presents the kurtosis (upper red curve) and skewness (lower blue curve) of $\rho_{ij}$.
		Both horizontal dashed straight lines indicate the locations of 3 and 0, which are the kurtosis and skewness of the Gaussian distribution, respectively.
		In these three panels,  the data points are plotted at the locations of the start dates of each time window.
	}
	\label{fig1}
\end{figure}

This paper first investigates the basic properties of the distributions of the correlation coefficients $\rho_{ij}$ in each time window
because the networks are based on them.
The data presented in the top panel of fig.~\ref{fig1} shows that the probability of anti-correlation $p\left(\rho_{ij} < 0\right)$ is as high as 0.43,
and the time windows with $p\left(\rho_{ij} < 0\right) > 0.1$ account for 31\% of the total windows.
From the middle panel, we observe that the minimum of $\rho_{ij}$ is as small as -0.92,
and the time windows where the minimum of $\rho_{ij}$ is smaller than -0.5 account for 90\% of the total windows.
For example, in the last time window,
the minimum of $\rho_{ij}$ is -0.81, and $p\left(\rho_{ij} < 0\right)$ is as high as 15\% (1,902,170 edges in the corresponding anti-correlation network).
These values mentioned above further illustrate the necessity and importance of studying anti-correlation networks.

To further investigate the shape of the distributions of correlation coefficients quantitatively,
the bottom panel of fig.~\ref{fig1} presents the kurtosis and skewness of the distributions in all 387 time windows.
This panel shows that the distributions are slightly platykurtic in the majority of time windows, but are extremely leptokurtic during the periods of market crashes.
This panel also shows that the distributions are negatively skewed in almost all time windows, and are more negatively skewed during periods of market crashes.
The dramatic changes in kurtosis and skewness during the periods of market crashes impact the topological structures of networks significantly.

\begin{figure}[H]
	\centering
	\subfigure{
		\includegraphics[width=0.495\linewidth]{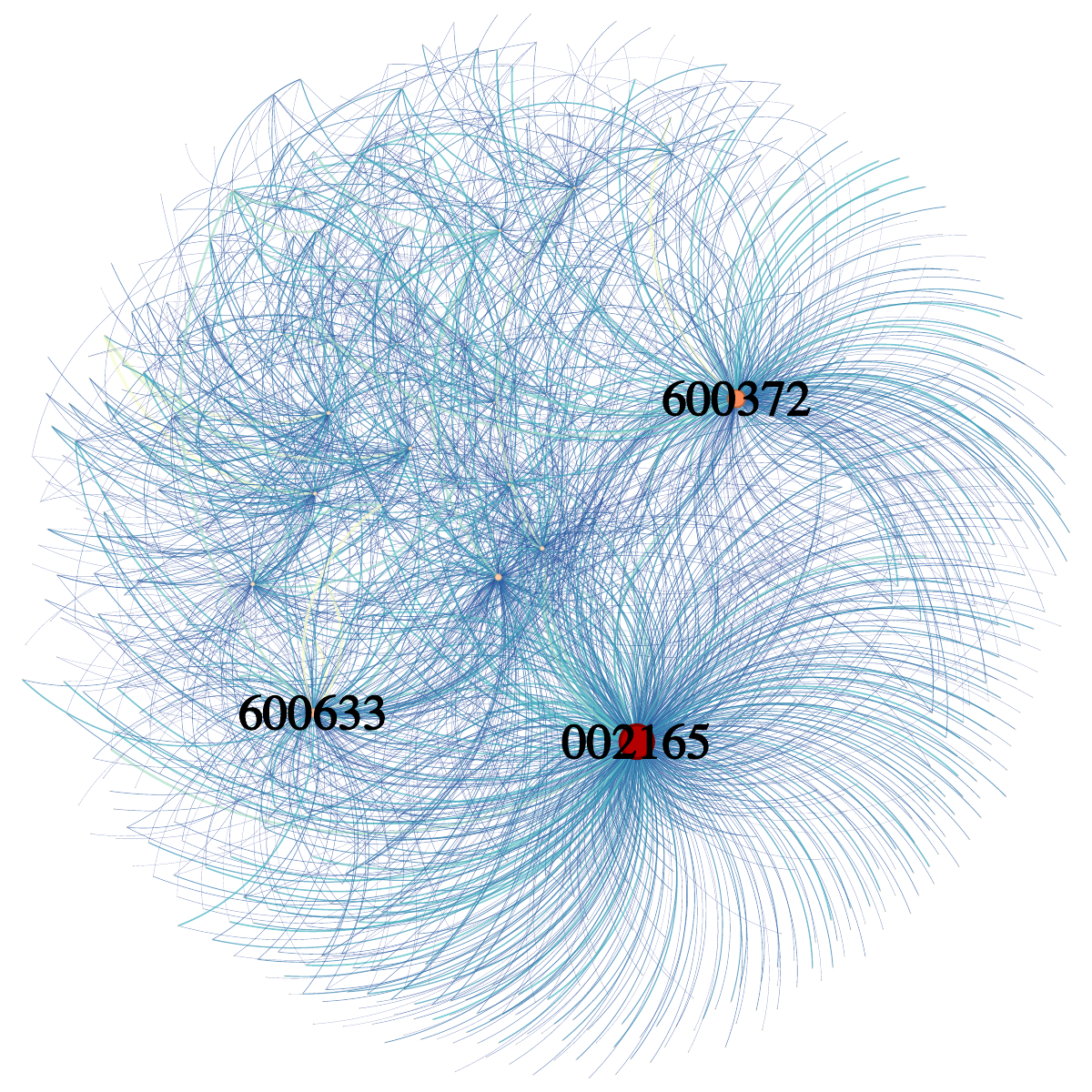}
	}
	\hspace{-6mm}
	\subfigure{
		\includegraphics[width=0.495\linewidth]{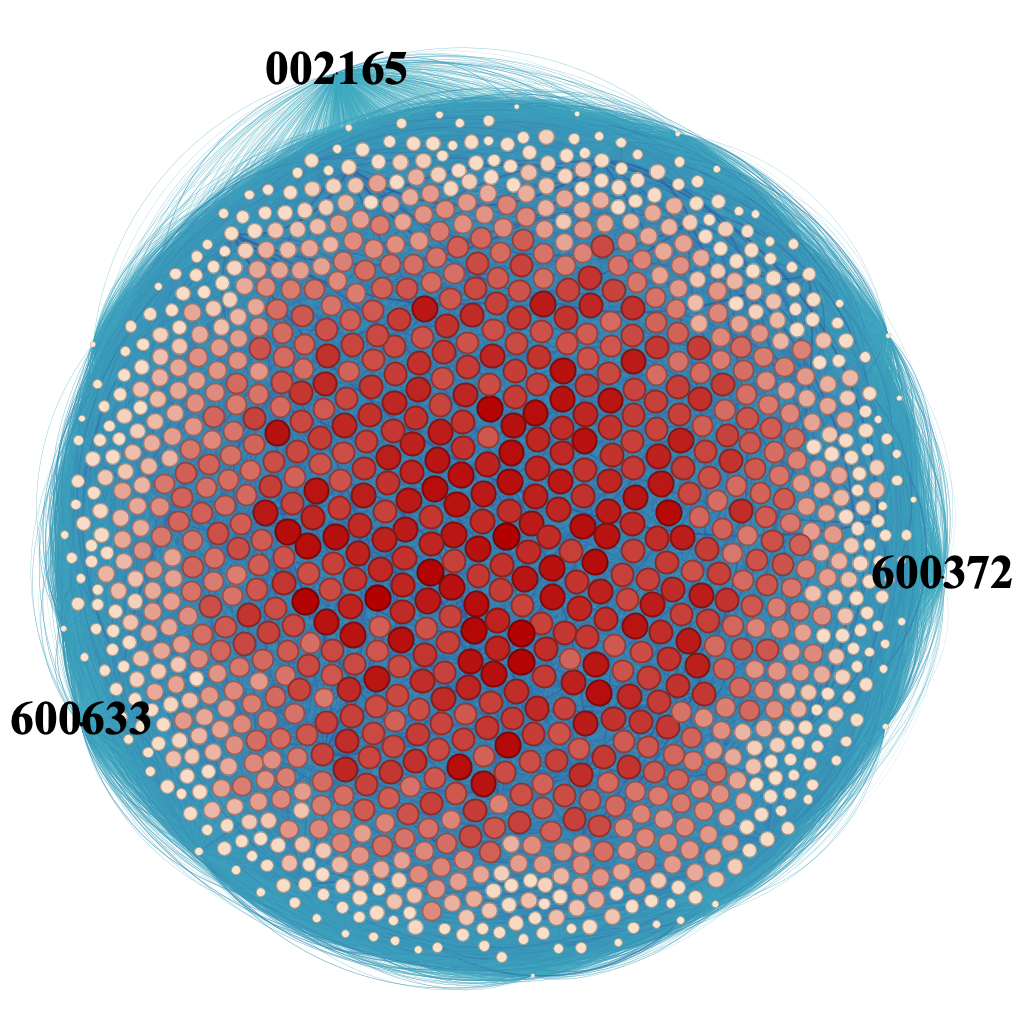}
	}
	\caption{The visualization of an anti-correlation network and a fully connected network.
	         The left panel shows a typical anti-correlation network with 994 nodes and 2,300 edges.
		This anti-correlation network is in the time window through Aug. 5, 2008 to Sept. 9, 2008, during which period the 2007-2008 global financial crisis was happening.
		In this panel, the nodes labeled by stock symbols are the top 3 stocks ranked by node's strength.
		The right panel shows the corresponding fully connected network with 1,124 nodes and 631,126 edges (see main text for details).
		In this network, the three tiny black nodes labeled by stock symbols are the top 3 stocks of the anti-correlation network shown in the left panel.
		In these two panels,
		a colored circle (node) represents a stock, whose color and size depend on its strength;
		a colored curve (edge) represents the correlation relationship between a pair of stocks linked by that curve, whose color and thickness depend on the corresponding correlation coefficient.
	}
	\label{fig2}
\end{figure}

To have an intuitive understanding of the structure of the anti-correlation network,
the left panel of fig.~\ref{fig2} shows a visualization of the anti-correlation network in the time window through Aug. 5, 2008 to Sept. 9, 2008,
during which period the 2007-2008 global financial crisis was happening.
This panel demonstrates that this anti-correlation network has a few extremely huge and critical nodes that are labeled by stock symbols.
For ease of comparison, the right panel shows the corresponding fully connected network in the time window through Aug. 5, 2008 to Sept. 9, 2008.
This network is constructed using a traditional method described in the introduction section, which cannot disentangle the positive and negative correlations.
Compared with the anti-correlation network,
in the fully connected network,
there are so many huge nodes that no critical nodes exist.

Fig.~\ref{fig2} illustrates that the anti-correlation network and the fully connected network have completely different structures.
The extremely huge and critical nodes in the anti-correlation network become extremely small and trivial nodes in the corresponding fully connected network.
This figure further demonstrates that the traditional method ignored anti-correlations.
This paper also checks the corresponding positive correlation network, which has a similar structure to the corresponding fully connected network.
This paper further visualizes the anti-correlation networks, positive correlation networks, and fully connected networks in all time windows,
and finds that the structures of these three types of networks have no significant temporal changes.

To quantitatively study the difference in the structure and property between the anti-correlation and positive correlation networks in all time windows,
the following text investigates fundamental measurements for both types of networks.

\begin{figure}[H]
	\centering
	\includegraphics[width=1.0\linewidth]{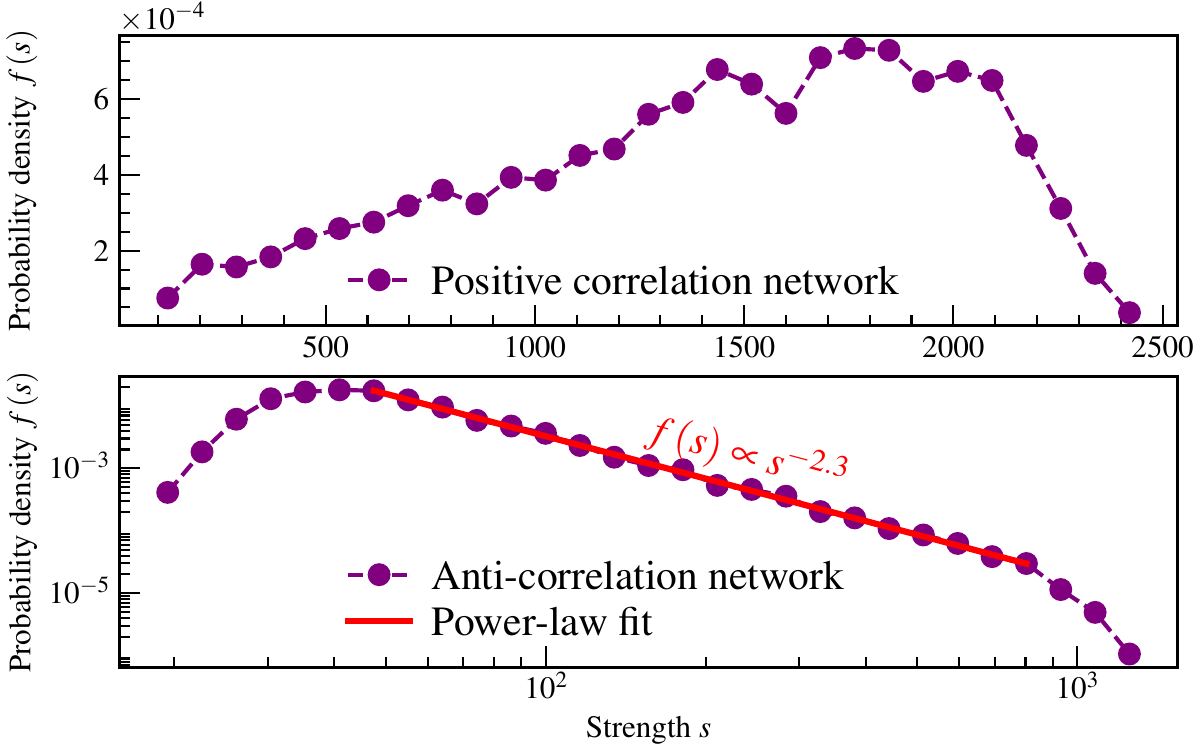}
	\caption{The probability density functions of strength for the anti-correlation and positive correlation networks.
		The upper and lower panels are for the positive correlation and anti-correlation networks in the last time window, respectively.
		The solid purple circles are the estimated probability density functions of strength $s$.
		The solid red straight line denotes the result of the power-law fit to data points using non-linear least squares.
	}
	\label{fig3}
\end{figure}

Fig.~\ref{fig3} presents strength distributions for the networks in the last time window.
It shows that
the empirical distribution of strength for the anti-correlation network is heavy-tailed.
This observation is further demonstrated by the power-law fit using non-linear least squares.
Such heavy-tailed distribution has been observed in many complex networks~{\cite{MST-2, MST-4, MST-6, Threshold-3, Threshold-5, ScalinginNetworks}}.
However, it is curious that the distribution for the positive correlation network presented here does not show the heavy-tailed behavior.

\begin{figure}[H]
	\centering
	\includegraphics[width=1.0\linewidth]{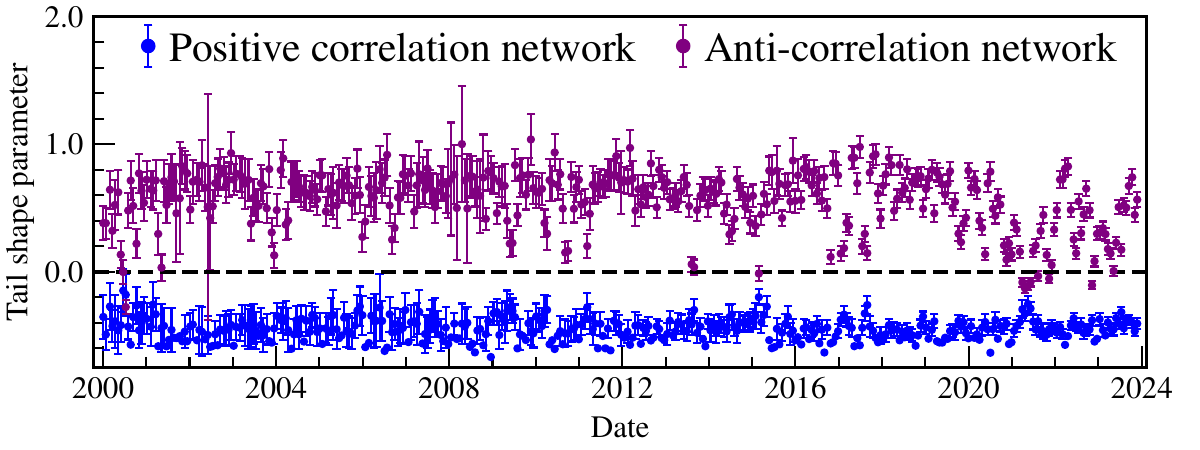}
	\caption{The estimations of the tail shape parameters of strength distributions for the anti-correlation and positive correlation networks.
		The shape parameters are estimated from the GPD fit to the tail data.
		The purple and blue circles are the point estimations, and the corresponding vertical lines with caps denote the 95\% confidence intervals.
		For the shape parameters below -0.5,
		only point estimations are shown because the confidence interval estimation is problematic.
		The markers are plotted at the locations of the start dates of each time window.
		For ease of comparison, the location of 0 is also denoted by the horizontal dashed line.
	}
	\label{fig4}
\end{figure}

To quantitatively examine the tail shapes of the strength distributions for the anti-correlation and positive correlation networks in all time windows,
this paper employs the Generalized Pareto Distribution (GPD) to fit tail data and then estimate the tail shape parameters using maximum likelihood estimation.
The GPD is widely used to estimate the tail shape parameter because it includes both cases of the thin and heavy-tailed.
The estimations are presented in fig.~\ref{fig4}.
It shows that the tail shape parameters for almost all anti-correlation networks are positive and smaller than 1,
and those for all positive correlation networks are negative.
A positive shape parameter indicates a heavy-tailed distribution.
This figure also shows that the confidence intervals in the last ten years have improved significantly because of the increasing statistics.

From the above studies,
this paper finds that almost all anti-correlation networks are scale-free in terms of strength,
while all positive correlation networks are not.
This finding is in agreement with the visualizations of networks.

The scale-free networks display a surprising degree of robustness against random failures of nodes
and extreme vulnerability from deliberate attacks by removing a few vital nodes~{\cite{scale-free-network}}.
Such attack vulnerability provides an important implication that
we can reduce market risks effectively by controlling a few stocks with high strength in the anti-correlation networks.
For positive correlation networks,
however,
it is hard to control market risks effectively because of the high interconnectedness and uniformity of the positive correlation networks, which are caused by the presence of massive highly correlated stocks.

In each time window,
this paper also observes the difference between the top 10 stocks ranked by node strength in the anti-correlation network and those in the positive correlation network.
The result indicates that there are no same stocks in the top 10 stocks between both types of networks in each time window.
It suggests that we should not only focus on the important stocks in the positive correlation networks as studied previously,
but also pay more attention to the key stocks in the anti-correlation networks studied in this paper.

To analyze the temporal changes of key stocks,
this paper constructs a collection of 3,870 stocks which come from the top 10 stocks of each anti-correlation network.
A similar collection with 3,870 stocks is also constructed for positive correlation networks.
The collection for anti-correlation networks has 1,711 distinct stocks,
and the collection for positive correlation networks has 1,765 distinct stocks.
In these two collections,
the maximum number of times a stock appears is 33 and 16,
and the number of stocks that appear more than ten times is 38 and 11,
for the anti-correlation and positive correlation networks, respectively.
This analysis indicates that the temporal changes of key stocks for the anti-correlation network are relatively infrequent compared with the positive correlation network.

\begin{figure*}[t]
	\centering
	\includegraphics[width=1.0\linewidth]{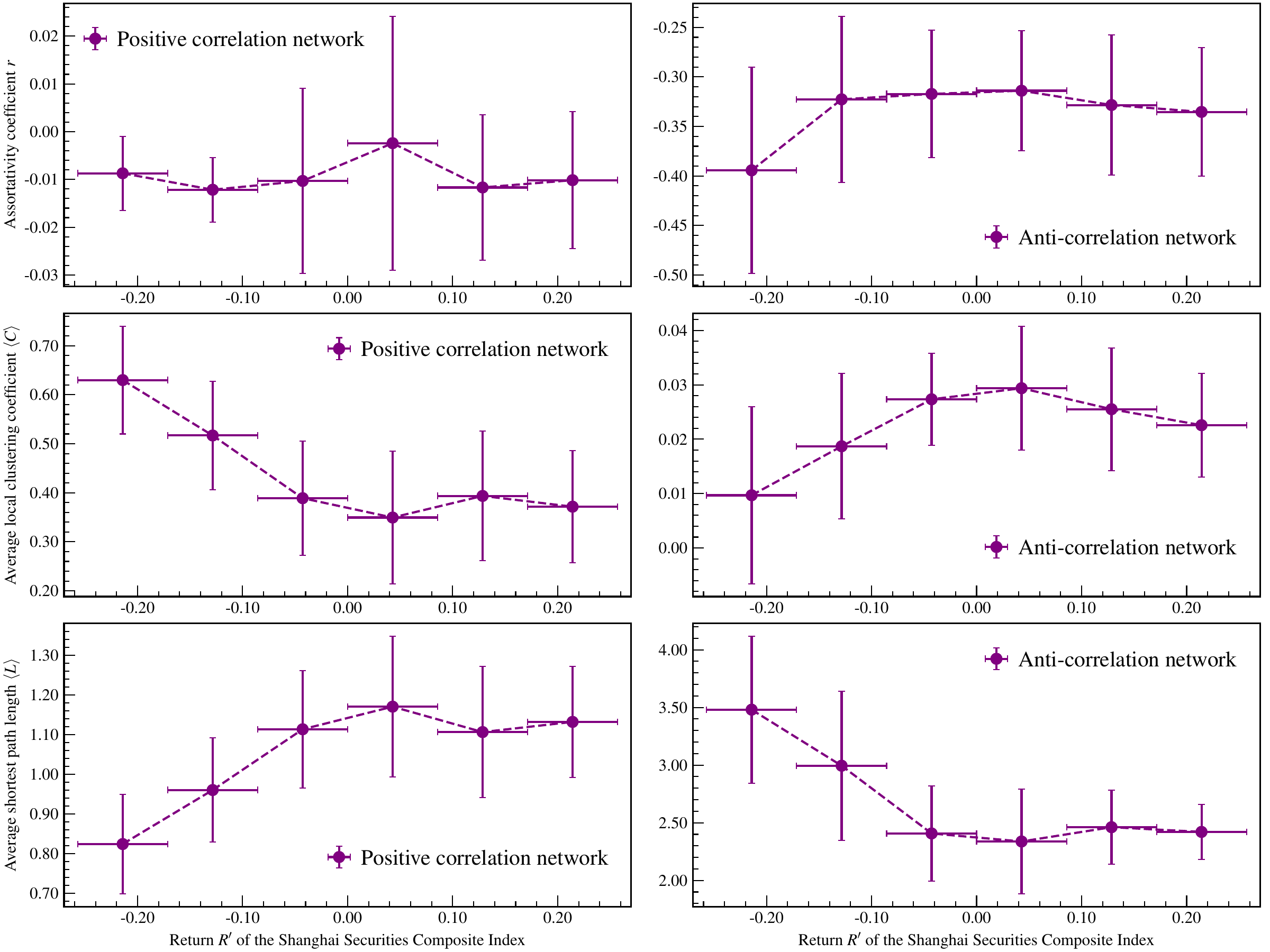}
	\caption{The assortativity coefficient (top panel), the average local clustering coefficient (middle panel), and the average shortest path length (bottom panel) as functions of the return of the Shanghai Securities Composite Index.
		The left and right panels are for the positive correlation and anti-correlation networks, respectively.
		The circles represent mean values in the specific ranges of return as shown by the horizontal lines with caps.
		The vertical lines with caps are the standard deviations.
	}
	\label{fig5}
\end{figure*}

Fig.~\ref{fig5} presents the data of the assortativity coefficient $r$ by strength, the average local clustering coefficient $\left< C \right>$,
and the average shortest path length $\left< L \right>$ for the positive correlation and anti-correlation networks.
To quantitatively investigate the effect of stock market fluctuations,
this figure also studies the relationships between these measurements and the return $R^{\prime}$ of the Shanghai Securities Composite Index.
The return $R^{\prime}$ in the $m$th time window is defined as $R^{\prime} = \ln C_{m} - \ln O_{m}$,
where $C_{m}$ and $O_{m}$ are the closure and open indices of the Shanghai Securities Composite Index in the $m$th time window, respectively.

The top panels of fig.~\ref{fig5} present the assortativity coefficients $r$.
The coefficients of positive correlation networks are close to 0.
Differently,
all anti-correlation networks behave significantly disassortative mixing.
This finding is in agreement with the visualizations of networks.
Compared with the positive correlation networks,
the feature of assortative mixing for the anti-correlation networks is more sensitive to market crashes.
The stock market decline ($R^{\prime} < 0$) has no significant effect on the assortativity coefficient $r$ for the positive correlation network.
However, the market crash seems to decrease that for the anti-correlation network.

The middle panels of fig.~\ref{fig5} present the average local clustering coefficients $\left< C \right>$.
The coefficients $\left< C \right>$ for the positive correlation networks are significantly larger than those for the anti-correlation networks.
These two panels demonstrate that the market crashes have opposite effects on the coefficients $\left< C \right>$ for the positive correlation and anti-correlation networks.
When $R^{\prime} < 0$,
the coefficient $\left< C \right>$ for the positive correlation network is a monotonic decreasing function of return $R^{\prime}$,
while $\left< C \right>$ for the anti-correlation network is a monotonic increasing function.

The significantly disassortative coefficients $r$ and extremely small average local clustering coefficients $\left< C \right>$ indicate that a lot of star-like structures exist in the anti-correlation networks.
The star-like structures are also observed in the visualizations of the anti-correlation networks.
In a star-like network, the hub nodes play a vital role in maintaining the network's structure and function.
This reminds us that we should pay more attention to the key stocks in the anti-correlation networks.

The bottom panels of fig.~\ref{fig5} present the average shortest path lengths $\left< L \right>$.
The $\left < L \right>$ for the positive correlation networks is smaller than that for the anti-correlation networks.
These two panels demonstrate that the market crashes have opposite effects on $\left< L \right>$ for these two types of networks.
In the time windows when market crashes happen,
$\left< L \right>$ significantly decreases and increases for the positive correlation and anti-correlation networks, respectively.
These significant changes in $\left< L \right>$ may be caused by the synchronization of the fluctuations of stock prices during periods of market crashes.
For a network,
a small $\left< L \right>$ indicates a strong ability of risk propagation in a financial system.
This implies that the anti-correlation network has an important role in stabilizing the stock market and in optimizing investment portfolios.

Fig.~\ref{fig5} demonstrates surprisingly that
the stock market decline ($R^{\prime} < 0$) has significant effects on these three measurements, but the stock market rise ($R^{\prime} > 0$) has no significant effects.
This asymmetry indicates that the decline event strengthens the collective behavior of investors,
but the rise event cannot strengthen that behavior.
This asymmetry mechanism needs to be studied further.

\section{Conclusions}
The correlation-based financial networks are intensively studied.
However,
previous studies ignored the importance of the anti-correlation.
Compared with the positive correlation,
the anti-correlation plays a more important role in understanding specific properties of a complex financial system.

To pay attention to the anti-correlation,
this paper is the first to consider the anti-correlation and positive correlation separately,
and accordingly construct the weighted anti-correlation and positive correlation networks among the 5,329 stocks listed in the Shanghai and Shenzhen stock exchanges.
To investigate the temporal evolution characteristics of the networks,
this paper uses the technique of sliding time window and then constructs 387 anti-correlation networks and 387 positive correlation networks during the 24 years from Jan. 1, 2000 to Dec. 31, 2023.

This work focuses on the differences in topological structures between the anti-correlation and positive correlation networks,
and systematically analyzes the most fundamental network's topological measurements:
The node's strength, the assortativity coefficient, the average local clustering coefficient, and the average shortest path length.
This paper finds some essential differences between both types of networks,
and concludes these findings as follows.
(1)
Almost all anti-correlation networks are scale-free in terms of strength, while all positive correlation networks are not;
the top 10 stocks ranked by the node strength and their temporal changes between the anti-correlation and positive correlation networks across all time windows are not the same.
(2)
The anti-correlation networks behave significantly disassortative mixing, while the assortativity coefficients of the positive correlation networks are close to 0;
the average local clustering coefficients for the anti-correlation networks are significantly smaller than those for the positive correlation networks;
the average shortest path lengths for the anti-correlation networks are larger than those for the positive correlation networks.
(3)
The stock market crash seems to decrease the assortativity coefficient of the anti-correlation network, while the stock market decline has no significant effect on the assortative mixing behavior of the positive correlation network;
the stock market decline has opposite effects on the anti-correlation and positive correlation networks in terms of both the average local clustering coefficient and the average shortest path length;
the stock market rise has no significant effects on these three topological measurements for both the anti-correlation and positive correlation networks.

These measurements reflect the functions of the anti-correlation and positive correlation networks.
Results indicate that these two types of networks play distinct roles in understanding complex financial systems, such as in risk contagion and stock market stabilization.
The anti-correlation network should receive more attention when investors optimize their portfolios and governments manage the market risks.
This paper also observes an asymmetry effect between the stock market decline and rise.
This asymmetry mechanism needs to be studied further.

The methodology proposed here has the potential to reveal significant differences
in the topological structure and dynamics of a complex financial system, stock behavior,  investment portfolios, and risk management,
offering insights that are not visible when all correlations are considered together.
More importantly,
this paper proposes a new direction for studying complex systems: the anti-correlation network.
It is well worth reexamining previous relevant studies using this new methodology.
%%%%%%%%%%%%%%%%%%%%%%%%%%%%%%%%%%%%%%%%%%%%%%%%%%%%%%%%%%%%%%%%%
%%%%%%%%%%%%%%%%%%%%%%%%%%%%%%%%%%%%%%%%%%%%%%%%%%%%%%%%%%%%%%%%%
%%%%%%%%%%%%%%%%%%%%%%%%%%%%%%%%%%%%%%%%%%%%%%%%%%%%%%%%%%%%%%%%%

\acknowledgements
This work was supported by the Humanities and Social Sciences Youth Foundation of the Ministry of Education of China [grant number 22YJCZH107];
the Shaanxi Science and Technology Department, P.R. China [grant number 2023-JC-QN-0093].

\bibliography{refs}

\end{document}